\documentclass[intlimits,twoside,a4paper]{article}
\usepackage{gensymb}
 \usepackage{amsmath,amssymb} 
 \usepackage{graphicx}      
 \usepackage{float}   
 \usepackage{bm}            
  \usepackage[T2A,T1]{fontenc} 
\usepackage[cp1251]{inputenc}
\usepackage{subfigure}
\usepackage{caption}

\usepackage{mhchem}
\usepackage[eqsecnum]{cmpj3}
\usepackage{bm}

\issue{2021}{24}{4}{43602}
\doinumber{10.5488/CMP.24.43602}

\title[Physical properties of \ce{{Cmc2}1-Si2P2X}: first-principles calculations]%
{Investigations on \ce{{Cmc2}1-Si2P2X} structures and physical properties by first-principles calculations%
}
\author[R. Yang, X. Gao, F. Wu, Q. Wei, M. Xue]{R. Yang\orcid{0000-0002-5962-5844}\refaddr{label1}\thanks{Corresponding author. E-mail: yrk18687@163.com}, 
        X. Gao\refaddr{label1}, F. Wu\refaddr{label1}, Q. Wei\refaddr{label1}, M. Xue\refaddr{label2}}
\addresses{
\addr{label1}{School of Physics and Optoelectronic Engineering, Xidian University, Xi'an, Shaanxi 710071, PR China}
\addr{label2} {College of Chemistry and Chemical Engineering, Guangxi University, Nanning, Guangxi 530004, PR China }
}

\date{Received October 09, 2020, in final form August 21, 2021}

\begin{document}

\maketitle

\begin{abstract}
The new structures, \ce{{Cmc2}1-Si2P2X} (X=S,~Se,~Te,~and~Po), are predicted, and their mechanical, electronic and optical properties are investigated with the density functional theory, by first principles calculations. The elastic constants of the four compounds are calculated by the stress-strain method. The calculations of the elastic stability criteria and phonon dispersion spectra imply that they are mechanically and dynamically stable at zero pressure. The mechanical parameters, such as shear moduli $G$, bulk moduli $B$, Young's moduli $E$ and Poisson's ratios $v$ are evaluated by the Voigt-Reuss-Hill approach.  The \ce{{Cmc2}1-Si2P2S}  has the largest hardness due to the largest Young's modulus in the four compounds, and it is a covalent crystal. The anisotropies of their mechanical properties are also analyzed. The band structures and densities of states, which are calculated by using HSE06, show that \ce{{Cmc2}1-Si2P2X} compounds are indirect bandgap semiconductors, and the values of the band gaps decrease with increasing atomic number from S, Se, Te, to Po. In addition, the longitudinal sound velocity and transverse sound velocity for \ce{{Cmc2}1-Si2P2X}  have been investigated. The dielectric constant, electron energy loss, refractive index, reflectivity, absorption and conductivity are analyzed to gain the optical properties of \ce{Si2P2X}.

\keywords \ce{{Cmc2}1-Si2P2X}  (X=S,~Se,~Te,~and~Po), first principles, mechanical properties, electronic properties, optical properties
%
\end{abstract}

\section{Introduction}

Silicon compounds, such as silicon dioxide, silicon nitride and silicon oxynitride, have aroused considerable interest in ceramic materials and electronic industry for their excellent physical and chemical properties. \ce{Si2N2O} is an engineering material that at temperatures up to $1600^{\circ}$C  still maintain excellent oxidation resistance, and high flexural strength without significant degradation at temperatures up to $1400^{\circ}$C~\cite{1,2}. The early works studied the ground state properties of orthorhombic-\ce{Si2N2O}(\ce{{Cmc2}1}) structure, i.e., electronic structure, elasticity, stability and optical properties~\cite{3,4,5,6,7}. Liu et al. researched the ground state, dielectric constants, and elastic stiffness of orthorhombic-\ce{Si2N2O} with various intrinsic point defects~\cite{8}. Zhen Zhu et al. proposed IV-VI compounds as isoelectronic counterparts to layered group V semiconductors, which have significantly broadened the scope of group IV semiconductors. The electronic band structure shows that \ce{SiSP2} is an indirect bandgap semiconductor~\cite{9}. Lv et al. researched dielectric properties, vibrational and ideal strength of \ce{Si2N2O} ceramic using the first principles~\cite{10}. 

In our previous works~\cite{11}, based on the similar properties of nitrogen and phosphorus, the structures of \ce{Si2P2O} and \ce{Ge2P2O} were derived, and their mechanical and electronic properties were studied under high pressure. The physical and chemical properties of group-VA (O, S, Se, Te, Po) are similar, according to the structure of \ce{Si2N2O ({Cmc2}1)}, the \ce{{Cmc2}1-Si2P2X} compounds are predicted. By consulting the literature, theoretical and experimental investigations on them have not been carried out up to now.

In this paper, the structure parameters, mechanical, electronic and optical properties of \ce{{Cmc2}1-Si2P2X} are calculated by using the density functional theory (DFT) method based on plane waves and pseudo-potentials. The stabilities of these structures have been proved through independent elastic constants and phonon dispersion spectra.

\section{ Calculation methods}
For \ce{{Cmc2}1-Si2P2X} (X=S, Se, Te, and Po), the structures are optimized and the properties are predicted using the DFT method with the generalized gradient approximations of Perdew-Burke-Ernzerhof (GGA-PBE), in the Cambridge Serial Total Energy Package (CASTEP) code~\cite{12,13}. CASTEP adopts the DFT plane-wave pseudo-potential method, which is used for first principles quantum mechanics calculations that can simulate a wide range of material properties including a structure at the atomic level, vibrational properties, electronic response properties, spectroscopic features etc. 

For \ce{{Cmc2}1-Si2P2X}, the energy cut-off values are set to 700 eV, which determines the number of plane waves, the $k$-point samplings with $9\times 9\times 8$ in the first irreducible Brillouin zone~\cite{14}. For the geometry optimization, the value of self-consistent field tolerance threshold is taken as  $5.0\times10^{-7}$~eV/atom. The convergence accuracy is set as ultrafine. The convergent value of total energy difference is less than $5.0\times10^{-6}$~eV/atom. The maximum force on the atom and the maximum stress are set to be 0.01~eV/\AA~and $0.02$~GPa. The phonon calculations are carried out within the harmonic approximation, using the linear response approach based on the density functional perturbation theory (DFPT) calculation, to determine its phonon dispersion spectra~\cite{15,16}. In addition, the hybrid Heyd-Scuseria-Ernzerhof (HSE06) functional is used in order to get more accurate electronic properties since GGA-PBE functional underestimates the bandgap energy of most semiconductors~\cite{17,18}.

The elastic constants are calculated using the stress-strain methods. The number of steps for each strain is set as 6 during the strain calculations, and the maximum strain amplitude is 0.003. In order to analyze the mechanical stabilities and calculate the mechanical moduli of \ce{{Cmc2}1-Si2P2X}, the elastic constants and other mechanical properties are calculated with GGA-PBE. For an orthorhombic crystal system, the elastic stability criteria are as follows:
\begin{equation}\begin{array}{l}
C_{i j}>0~(i=0,1,2,3,4,5,6),~ C_{11}+C_{12}+C_{33}+2\left(C_{12}+C_{13}+C_{23}\right)>0, \\
C_{11}+C_{22}-2 C_{12}>0,~ C_{11}+C_{33}-2 C_{13}>0,~ C_{11}+C_{33}-2 C_{23}>0.
\end{array}\end{equation}

Based on the nine independent elastic constants, for \ce{{Cmc2}1-Si2P2X} structures, the shear modulus~$(G)$, bulk modulus $(B)$, Poisson's ratio $(v)$ and Young's modulus $(E)$ are obtained with the Voigt-Reuss-Hill approximation~\cite{19}. For orthorhombic lattices, the moduli of Reuss shear $(G_{\rm R})$ and the Voigt shear~$(G_{\rm V})$ are expressed as follows~\cite{20}
\begin{equation}
G_{\rm R}=\frac{15}{4\left(S_{11}+S_{22}+S_{33}\right)-4\left(S_{12}+S_{13}+S_{23}\right)+3\left(S_{44}+S_{55}+S_{66}\right)},
\end{equation}
\begin{equation}
G_{\rm V}=\frac{1}{15}\left(C_{11}+C_{22}+C_{33}-C_{12}-C_{13}-C_{23}\right)+\frac{1}{5}\left(C_{44}+C_{55}+C_{66}\right).
\end{equation}
The Voigt bulk modulus ($B_{\rm V}$) and the Reuss bulk modulus ($B_{\rm R}$) can be given by 
\begin{equation}
B_{\rm V}=\frac{1}{9}\left(C_{11}+C_{22}+C_{33}\right)+\frac{2}{9}\left(C_{12}+C_{13}+C_{23}\right), 
\end{equation}
\begin{equation}
B_{\rm R}=\frac{1}{\left(S_{11}+S_{22}+S_{33}\right)+2\left(S_{12}+S_{13}+S_{23}\right)}.
\end{equation}
The relationship between $C_{ij}$ and $S_{ij}$ is given by
\begin{equation}
\left[S_{i j}\right]=[C]^{-1}_{i j}.
\end{equation}
The actual polycrystalline moduli can be calculated by the Voigt and Reuss moduli, they are shown as~\cite{21}
\begin{equation}
B=\frac{1}{2}\left(B_{\rm V}+B_{\rm R}\right),
\end{equation}
\begin{equation}
G=\frac{1}{2}\left(G_{\rm V}+G_{\rm R}\right).
\end{equation}

Young's modulus, $E$, is the ratio of the stress and strain, which represents the stiffness property of a material. Poisson's ratio, $v$, can offer further information about the bonding force feature. The larger Poisson's ratio reflects a better plasticity. Young's modulus and Poisson's ratio can be expressed by~\cite{22,23}
\begin{equation}
E=\frac{9BG}{3B+G}, 
\end{equation}
\begin{equation}
v=\frac{3B-2G}{2(3B+G)}.
\end{equation}

The anisotropy of a crystal material reflects the different arrangement of atoms and different bonding properties in different directions, which affect the physical and chemical properties of materials in different directions. The percent anisotropy ($A_B$ and $A_G$ for bulk and shear moduli) and the anisotropic index $A^U$ describe the elastic anisotropy of a crystal material~\cite{24}. The equations are as follows~\cite{25,26}
\begin{equation}
A^{U}=5 \frac{G_{\rm V}}{G_{\rm R}}+\frac{B_{\rm V}}{B_{\rm R}}-6 \geqslant 0, 
\end{equation}
\begin{equation}
A_{B}=\frac{B_{\rm V}-B_{\rm R}}{B_{\rm V}+B_{\rm R}}, 
\end{equation}
\begin{equation}
A_{G}=\frac{G_{\rm V}-G_{\rm R}}{G_{\rm V}+G_{\rm R}}.
\end{equation}
The shear anisotropic factors $A_1$, $A_2$, $A_3$ are given by:
\begin{equation}
\text { for the }(100) \text { plane }  A_{1}=\frac{4 C_{44}}{C_{11}+C_{33}-2 C_{13}},
\end{equation}
\begin{equation}
\text { for the }(010) \text { plane }  A_{2}=\frac{4 C_{55}}{C_{22}+C_{33}-2 C_{23}},
\end{equation}
\begin{equation}
\text { for the }(010) \text { plane }  A_{3}=\frac{4 C_{66}}{C_{11}+C_{22}-2 C_{12}}.
\end{equation}

The three-dimensional (3D) graph of Young's modulus can more effectively describe the elastic anisotropy. The Young's modulus equation for orthorhombic crystal is expressed by
\begin{equation}
\frac{1}{E}=l_{1}^{4} s_{11}+l_{2}^{4} s_{22}+l_{3}^{4} s_{33}+l_{1}^{2} l_{2}^{2}\left(2 s_{12}+s_{66}\right)+l_{2}^{2} l_{3}^{2}\left(2 s_{23}+s_{44}\right)+l_{1}^{2} l_{3}^{2}\left(2 s_{13}+s_{55}\right).
\end{equation}

Once we have calculated the Young's modulus $E$, the bulk modulus $B$ and shear modulus $G$, the longitudinal, transverse and average sound velocities are obtained in the materials. The averaged sound velocity $v_m$ is approximately predicted by the following equation~\cite{27}:
\begin{equation}
v_{m}=\left[\frac{1}{3}\left(\frac{2}{v_{t}^{3}}+\frac{1}{v_{l}^{3}}\right)\right]^{-1 / 3},
\end{equation}
where $v_l$ and $v_t$ are longitudinal and transverse sound velocity, which can be obtained from the bulk modulus ($B$) and shear modulus ($G$), by the use of the Navier's equation~\cite{28,29}:

\begin{equation}
v_{1}=\left(\frac{B+\frac{4}{3} G}{\rho}\right)^{{1}/{2}},
\end{equation}

\begin{equation}
v_{t}=\left(\frac{G}{\rho}\right)^{{1}/{2}}.
\end{equation}

In the calculations of the optical properties, the norm-conserving pseudo-potentials were applied to treat the ion-electron interactions~\cite{30}. The optical properties of systems are evaluated by the complex dielectric function, which is dependent on frequency and is as follows
\begin{equation}
\varepsilon(\omega)=\varepsilon_{1}(\omega)+\ri \varepsilon_{2}(\omega).
\end{equation}
The imaginary part $\varepsilon_2(\omega)$ of the dielectric function could be calculated from the momentum matrix elements between the occupied and unoccupied wave functions, it is shown as~\cite{31}
\begin{equation}
\varepsilon_{2}=\frac{V e^{2}}{2 \piup \hbar m^{2} \omega^{2}} \int \rd^{3} k \sum\left|\left\langle k n|p| k n^{\prime}\right\rangle\right|^{2} f(k n) \left[1-f\left(k n^{\prime}\right)\right] \delta\left(E_{k n}-E_{k r^{\prime}}-\hbar \omega\right),
\end{equation}
where $V$ is the volume of unit cell, $p$ expresses the momentum operator, $e$ is the charge of a free electron, $f(kn)$ stands for the Fermi distribution function, $|k n\rangle$  represents a crystal wave function, and $\hbar\omega$ is the incident photon energy. 

The complex dielectric function describes the optical properties of a medium for different photon energies. The real part of dielectric function is in relation to the electric polarization characteristics of the materials. The peak value of the real part is related to the electron excitation. The real part  $\varepsilon_1(\omega)$ of the dielectric function can be derived from the imaginary part $\varepsilon_2(\omega)$ by Kramers-Kroning relationship~\cite{32}:
\begin{equation}
\varepsilon_{1}(\omega)=1+\frac{2}{\piup} M \int_{0}^{\infty} \frac{\varepsilon_{2}\left(\omega^{\prime}\right) \omega^{\prime}}{\omega^{\prime 2}-\omega^{2}} \rd \omega^{\prime},
\end{equation}
where $M$ is the principal value of the integral. CASTEP provides two main approaches to optical properties calculations; one is based on standard DFT Kohn-Sham orbitals and the other one uses time-dependent DFT (TD-DFT) theory. In this paper, the Kohn-Sham DFT is used. The DFT method uses the independent particle approximation (IPA) for the calculation of $\varepsilon_{2}(\omega)$ based on the ground-state quantities~\cite{33}. The optical properties are analyzed using PBE-GGA.

Furthermore, other properties of the materials can be calculated by the dielectric function, including the optical conductivity, absorption coefficient, energy-loss spectrum and optical reflectivity. They are given as follows~\cite{34,35,36}
\begin{equation}
\sigma(\omega)=\frac{\ri \omega}{4 \piup}[1-\varepsilon(\omega)],
\end{equation}
\begin{equation}
I(\omega)=\frac{2 \omega k(\omega)}{c},
\end{equation}
\begin{equation}
R(\omega)=\frac{(n-1)^{2}+k^{2}}{(n+1)^{2}+k^{2}},
\end{equation}
\begin{equation}
L(\omega)=\operatorname{Im}\left[\frac{-1}{\varepsilon(\omega)}\right]=\frac{\varepsilon_{2}(\omega)}{\varepsilon_{1}(\omega)+\varepsilon_{2}(\omega)}.
\end{equation}
Where $k(\omega)$ means the extinction coefficient. It can be obtained from $\varepsilon_{1}(\omega)$ and $\varepsilon_{2}(\omega)$, which can be expressed as follows~\cite{37}
\begin{equation}
k(\omega)=\left[\frac{\sqrt{\varepsilon_{1}^{2}(\omega)+\varepsilon_{2}^{2}(\omega)}-\varepsilon_{1}(\omega)}{2}\right]^{{1}/{2}}.
\end{equation}
\section{ Results and discussion}
\subsection{Structure properties}
\begin{figure}[!t]
\centerline{\includegraphics[width=0.80\textwidth]{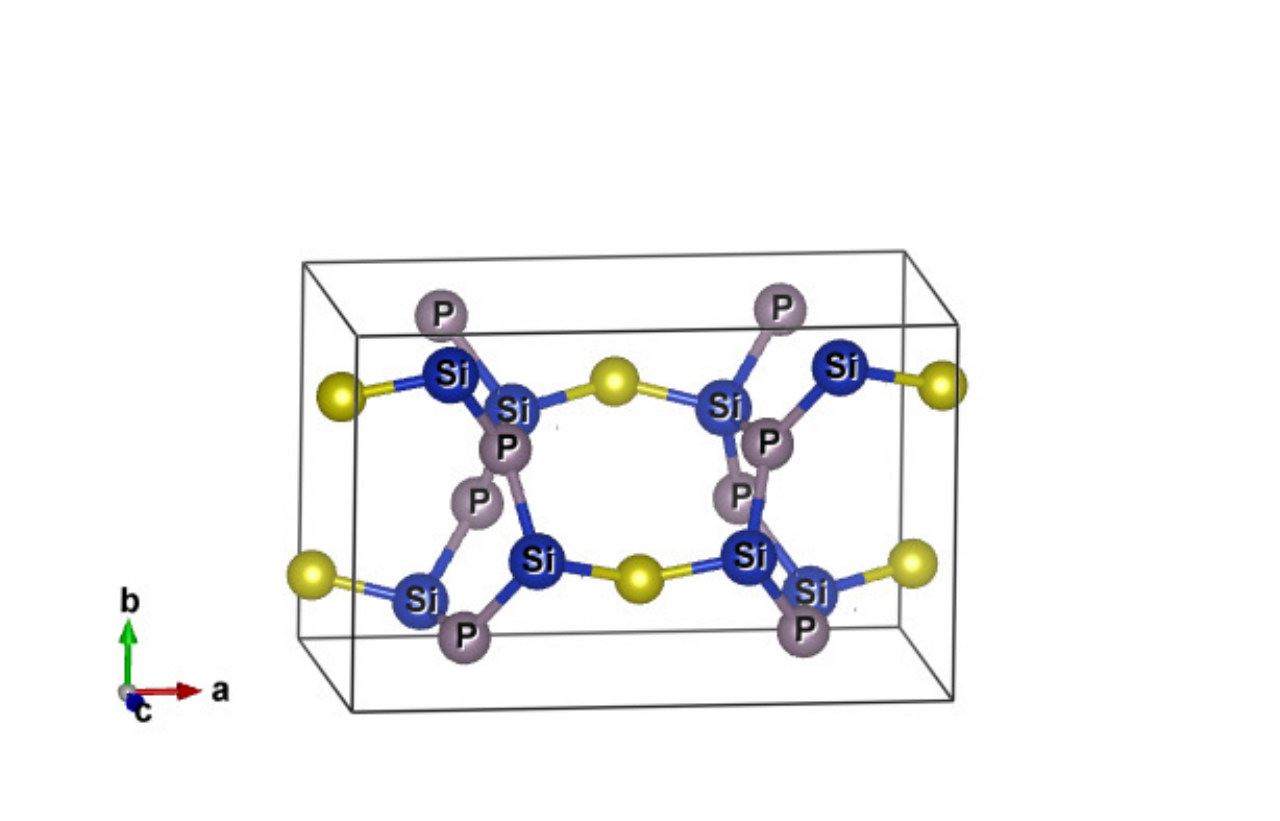}}
\caption{(Colour online) Crystal structures of orthorhombic-$\mathrm{Si_2}\mathrm{P_2X}$ (\ce{{Cmc2}1}, No.~36). The yellow balls represent X (S,~Se,~Te,~Po) atoms.} \label{fig-smp1}
\end{figure}

The compounds \ce{Si2P2X} have an orthorhombic structure with \ce{{Cmc2}1} space group (No.~36). The crystal structure is shown at 0~GPa as figure \ref{fig-smp1}, which is a centro-symmetric structure $(0, 0, 0)$ with 10 atoms per unit cell. On the basis of \ce{Si2P2O}, the lattice parameters and volumes of \ce{Si2P2S}, \ce{Si2P2Se}, \ce{Si2P2Te} and \ce{Si2P2Po} are predicted and presented in table \ref{tbl-smp1}. The density values of  \ce{{Cmc2}1-Si2P2X} are $2.43, 3.06, 3.47$, and $4.56$~g/cm$^3$, respectively. %

\begin{table}[htb]
\caption{Lattice constants $a$, $b$, $c$ (\AA) and volume (\AA$^3$) of the \ce{{Cmc2}1-Si2P2X}.}
\label{tbl-smp1}
\vspace{2ex}
\begin{center}
\renewcommand{\arraystretch}{0}
\begin{tabular}{|c||c|c|c|c|}
\hline 
Structure & $\boldsymbol{a}$ & $\boldsymbol{b}$ & $\boldsymbol{c}$ & $\boldsymbol{V}$\strut \\
\hline
\rule{0pt}{2pt}&&&&\\
\hline
\rule{0pt}{2pt}&&&&\\
$\mathrm{Si}_{2} \mathrm{P}_{2} \mathrm{S}$  & 10.73 & 6.39 & 5.97 & 409.23\strut \\
\cline{1-5}
\hline $\mathrm{Si}_{2} \mathrm{P}_{2} \mathrm{Se}$ & 11.03 & 6.45 & 5.99 & 426.91\strut \\
\cline{1-5}
\hline $\mathrm{Si}_{2} \mathrm{P}_{2} \mathrm{Te}$ & 11.89 & 6.50 & 6.06 & 469.74\strut \\
\cline{1-5}
\hline $\mathrm{Si}_{2} \mathrm{P}_{2} \mathrm{Po}$ & 12.23 & 6.46 & 6.02 & 475.68\strut \\
\hline
\end{tabular}
\renewcommand{\arraystretch}{1}
\end{center}
\end{table}

\subsection{Elastic properties and stability analysis}
The elastic properties of solids contain important information about their mechanical properties. For  \ce{{Cmc2}1-Si2P2X}, the nine independent elastic constants are calculated at 0~GPa and are shown in table~\ref{tbl-smp2}. 
\begin{table}[H]
\caption{Calculated elastic constants $C_{ij}$ of \ce{{Cmc2}1-Si2P2X} at 0 GPa.}
\label{tbl-smp2}
\vspace{2ex}
\begin{center}
\renewcommand{\arraystretch}{0}
\begin{tabular}{|c||c|c|c|c|c|c|c|c|c|}
\hline
Species&$C_{11}$&$C_{12}$&$C_{13}$&$C_{22}$&$C_{23}$&$C_{33}$&$C_{44}$&$C_{55}$&$C_{66}$\strut\\
\hline
\rule{0pt}{2pt}&&&&&&&&&\\
\hline
\cline{1-10}
      $\mathrm{Si}_{2} \mathrm{P}_{2} \mathrm{S}$ & 100.6&  18.7&  13.7&  75.0&  3.3&  133.2&  36.7&  34.5& 29.9\strut\\
\cline{1-10}
       $\mathrm{Si}_{2} \mathrm{P}_{2} \mathrm{Se}$ & 82.3&  24.7&  13.1&  82.8&  2.7&  134.7&  34.9&  27.2& 15.2\strut\\
\cline{1-10}
      $\mathrm{Si}_{2} \mathrm{P}_{2} \mathrm{Te}$ & 52.5&  22.5&  26.8&  71.7&  7.1&  127.7&  28.8&  29.7& 20.3\strut\\
\cline{1-10}
     $\mathrm{Si}_{2} \mathrm{P}_{2} \mathrm{Po}$ & 57.6&  29.0&  35.7&  74.4&  9.5& 130.3&  25.4& 37.9& 27.3\strut\\
\hline
\end{tabular}
\renewcommand{\arraystretch}{1}
\end{center}
\end{table}

\begin{figure}[H]
\centering
\subfigure{
\includegraphics[width=6.1cm,height=6.1cm]{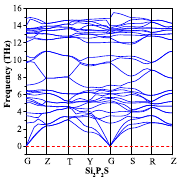}
}
\quad
\subfigure{
\includegraphics[width=6.1cm,height=6.1cm]{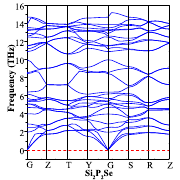}
}
\quad
\subfigure{
\includegraphics[width=6.1cm,height=6.1cm]{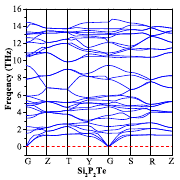}
}
\quad
\subfigure{
\includegraphics[width=6.1cm,height=6.1cm]{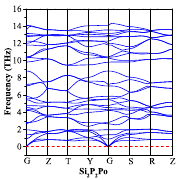}
}
\caption{(Colour online) The phonon dispersion spectra for \ce{{Cmc2}1-Si2P2X} at 0 GPa.} \label{fig-smp2}
\end{figure}

The \ce{{Cmc2}1-Si2P2X} structures are all mechanically stable due to satisfying the elastic stability criteria. To ensure the stability of  \ce{{Cmc2}1-Si2P2X}, the phonon spectra are calculated. The phonon dispersion spectra under 0~GPa are displayed in figure \ref{fig-smp2}. The calculated phonon spectra have no imaginary phonon frequencies in the whole Brillouin zone (BZ). It is proved that the four structures are dynamically stable~\cite{38,39}.

For the four compounds, the $C_{33}$ is larger than the values of $C_{11}$ and $C_{22}$. It is implied that the mechanical strength in the [001] direction is higher than that in the [100] and [010] directions. In addition, $C_{44}$, $C_{55}$ and $C_{66}$ values express the shear moduli in the (100), (010) and (001) crystal planes, respectively. For \ce{Si2P2S} and \ce{Si2P2Se}, the values for $C_{44}$ are larger than $C_{55}$ and $C_{66}$. It means that the shear modulus at (100) crystal plane is larger than that of (010) and (001) crystal planes. The $C_{55}$ values of \ce{Si2P2Te} and \ce{Si2P2Po} are larger than $C_{44}$ and $C_{66}$. Therefore, the shear modulus at (010) crystal plane is larger than that of (100) and (001). 

Based on the nine independent elastic constants, the shear modulus ($G$), bulk modulus ($B$), Poisson's ratio ($v$) and Young's modulus ($E$) of the \ce{Si2P2X} structures are calculated and the results are shown in table \ref{tbl-smp3}.

The bulk modulus $(B)$ and the shear modulus $(G)$ express the resistance to the volume change by load pressure and the resistance to deformation by shear stress, respectively. By calculations, the results show that the bulk modulus of the \ce{{Cmc2}1-Si2P2Po} is the largest among \ce{{Cmc2}1-Si2P2X}, which shows that it has a lower compressibility. The shear modulus of \ce{{Cmc2}1-Si2P2S} is higher than the others. Hence, it has a better capability of resisting to the shape change. The \ce{{Cmc2}1-Si2P2S} has the largest stiffness due to its largest Young's modulus in the four compounds. 

\begin{table}[htb]
\caption{The bulk moduli ($B_{\rm V}, B_{\rm R}$ and $B$), shear moduli ($G_{\rm V}, G_{\rm R}$ and $G$), Young's moduli ($E$), Poisson's ratios ($v$), and $B/G$.}
\label{tbl-smp3}
\vspace{2ex}
\begin{center}
\renewcommand{\arraystretch}{0}
\begin{tabular}{|c||c|c|c|c|c|c|c|c|c|}
\hline
Species&$B_{\rm V}$&$B_{\rm R}$&$G_{\rm V}$&$G_{\rm R}$&$B$&$G$&$E$&$v$&$B/G$\strut\\
\hline
\rule{0pt}{2pt}&&&&&&&&&\\
\hline
\cline{1-10}
      $\mathrm{Si}_{2} \mathrm{P}_{2} \mathrm{S}$ & 42.25&  40.37&  38.43&  36.38&  41.32&  37.40&  86.19&  0.15& 1.10\strut\\
\cline{1-10}
       $\mathrm{Si}_{2} \mathrm{P}_{2} \mathrm{Se}$ &42.34	&41.48&32.74&27.20&41.91&29.97&72.60&	0.21&	1.39\strut\\
\cline{1-10}
      $\mathrm{Si}_{2} \mathrm{P}_{2} \mathrm{Te}$ & 40.54&37.35&28.81&25.15&38.94&26.98&65.75&0.22&1.44\strut\\
\cline{1-10}
     $\mathrm{Si}_{2} \mathrm{P}_{2} \mathrm{Po}$ & 45.65&42.78&30.66&26.12&44.22&28.39&70.15&0.23&1.55\strut\\
\hline
\end{tabular}
\renewcommand{\arraystretch}{1}
\end{center}
\end{table}

The Poisson's ratio of ductile materials is larger than~0.33, while the $v$ for brittle material is less than~0.33~\cite{40}. Therefore, the results indicate that the compounds \ce{{Cmc2}1-Si2P2X} are all brittle materials. The Poisson's ratio of a strong covalent crystal is often below 0.15. For \ce{{Cmc2}1-Si2P2S}, the calculated Poisson's ratio shows that it is a covalent crystal. The value of $B/G$ means the brittleness or ductility of materials, whose the judgment value is 1.75~\cite{41}. Above 1.75, the material shows toughness; otherwise it shows brittleness. Obviously, the \ce{{Cmc2}1-Si2P2S} is more brittle than the others. 

\begin{table}[htb]
\caption{The calculated anisotropic index ($A^U$), shear anisotropic factors ($A_1$, $A_2$, $A_3$), and percent anisotropy ($A_B$ and $A_G$) of \ce{{Cmc2}1-Si2P2X}.}
\label{tbl-smp4}
\vspace{2ex}
\begin{center}
\renewcommand{\arraystretch}{0}
\begin{tabular}{|c||c|c|c|c|c|c|}
\hline
&$A^U$&$A_B$&$A_G$&$A_1$&$A_2$&$A_3$\strut\\
\hline
\rule{0pt}{2pt}&&&&&&\\
\hline
\cline{1-7}
      $\mathrm{Si}_{2} \mathrm{P}_{2} \mathrm{S}$ & 0.33&  0.02&  0.03&  0.71&  0.68&  0.86\strut\\
\cline{1-7}
       $\mathrm{Si}_{2} \mathrm{P}_{2} \mathrm{Se}$ & 1.04&  0.01&  0.09& 0.73&  0.51&  0.52\strut\\
\cline{1-7}
      $\mathrm{Si}_{2} \mathrm{P}_{2} \mathrm{Te}$ & 0.81& 0.04&  0.07&  0.91&  0.64&  1.02\strut\\
\cline{1-7}
     $\mathrm{Si}_{2} \mathrm{P}_{2} \mathrm{Po}$ & 0.93&  0.03&  0.08&  0.87&  0.81& 1.47\strut\\
\hline
\end{tabular}
\renewcommand{\arraystretch}{1}
\end{center}
\end{table}

Crystal anisotropy can be expressed by different properties, such as the hardness, optical properties, elastic modulus and thermal properties and so on. The elastic anisotropy plays an important role in many mechanical and physical properties, such as unusual phonon modes, phase transformations, anisotropic plastic deformation, elastic instability, etc.~\cite{42}. When the values of $A^U$, $A_B$ and $A_G$ are 0, then the crystal is isotropic. The values of the $A_B$ of the four crystals are close to 0, which means that their bulk moduli are nearly isotropic. The deviation of $A^U$ from zero indicates the extent of crystal anisotropy and accounts for both the shear and the bulk contributions~\cite{43,44}. Thus, $A^U$ represents a universal measure to quantify the crystal elastic anisotropy. The anisotropic parameters of the crystal materials are listed in table~\ref{tbl-smp4}. Based on the calculations, the elastic anisotropic index of \ce{{Cmc2}1-Si2P2Se} has a larger value than the other compounds, suggesting that the \ce{{Cmc2}1-Si2P2Se} shows a stronger anisotropy than the others.

\begin{table}[htb]
\caption{ Calculated density $\rho$  (in g/cm$^3$), longitudinal, transverse and average sound velocity ($v_l$, $v_t$, $v_m$, respectively, in m/s) for  \ce{{Cmc2}1-Si2P2X}.}
\label{tbl-smp5}
\vspace{2ex}
\begin{center}
\renewcommand{\arraystretch}{0}
\begin{tabular}{|c||c|c|c|c|}
\hline
&$\rho$&$v_l$&$v_t$&$v_m$\strut\\
\hline
\rule{0pt}{2pt}&&&&\\
\hline
\cline{1-5}
      $\mathrm{Si}_{2} \mathrm{P}_{2} \mathrm{S}$ & 2.43&  6125&  3923&  4309\strut\\
\cline{1-5}
       $\mathrm{Si}_{2} \mathrm{P}_{2} \mathrm{Se}$ & 3.06&  5172&  3129& 3459\strut\\
\cline{1-5}
      $\mathrm{Si}_{2} \mathrm{P}_{2} \mathrm{Te}$ & 3.47&  4646&  2788&  3084\strut\\
\cline{1-5}
     $\mathrm{Si}_{2} \mathrm{P}_{2} \mathrm{Po}$ & 4.56&  4242&  2495&  2765\strut\\
\hline
\end{tabular}
\renewcommand{\arraystretch}{1}
\end{center}
\end{table}

\begin{figure}[!t]
\centering
\subfigure{
\includegraphics[width=7cm,height=6.85cm]{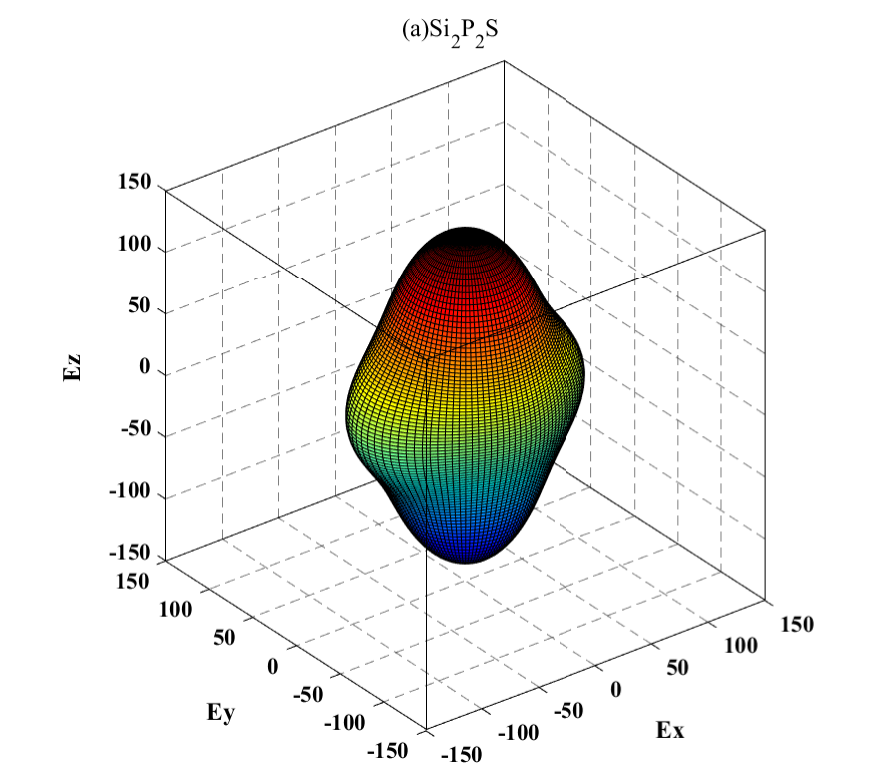}
}
\quad
\subfigure{
\includegraphics[width=7cm,height=6.85cm]{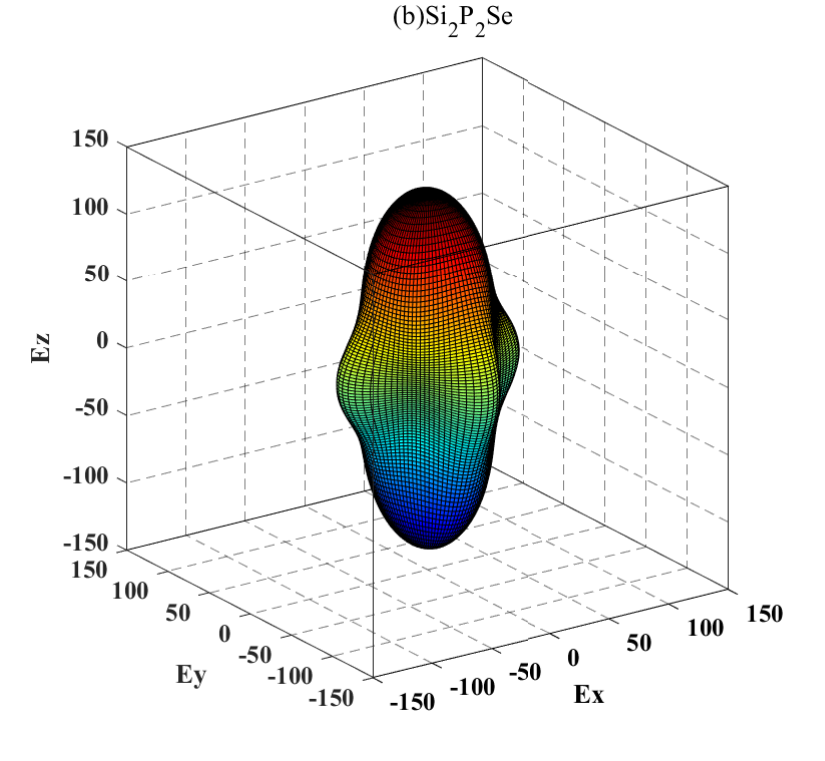}
}
\quad
\subfigure{
\includegraphics[width=7.5cm,height=6.85cm]{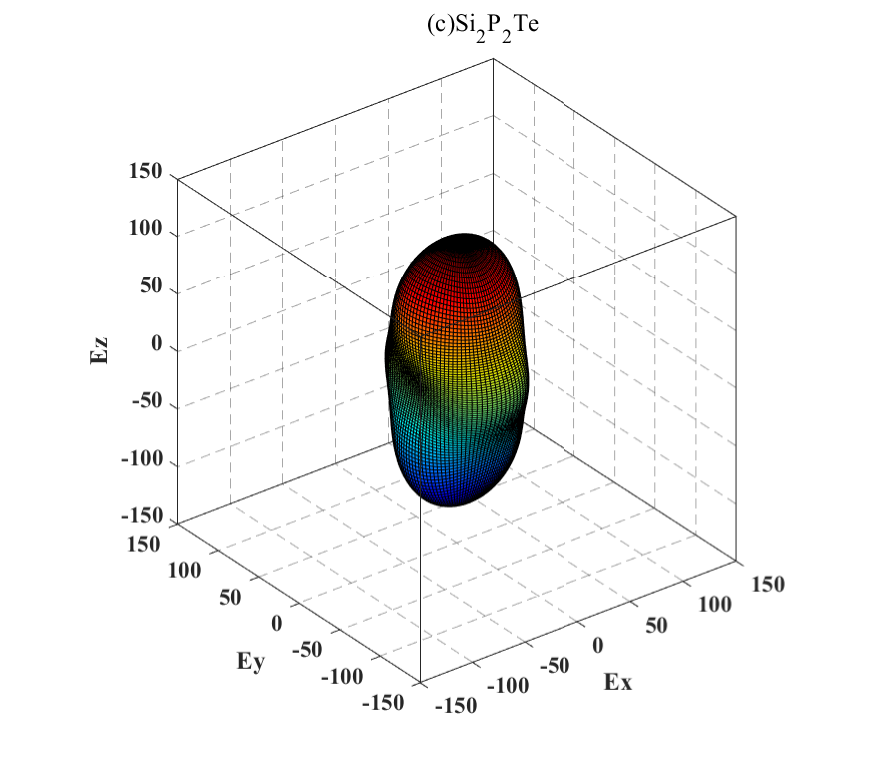}
}
\quad
\subfigure{
\includegraphics[width=6.5cm,height=6.85cm]{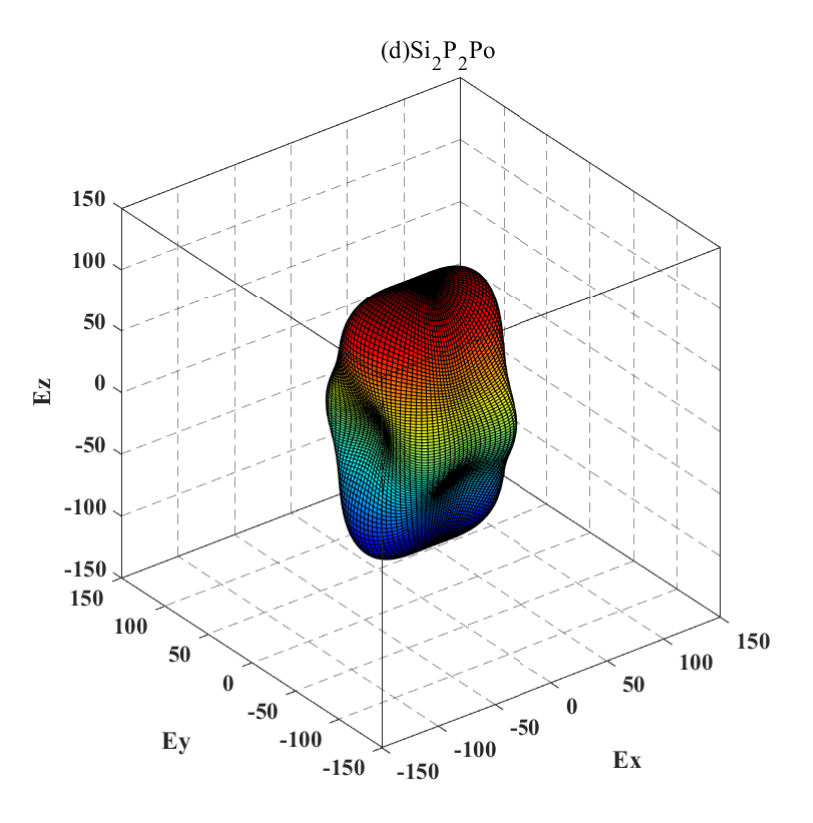}
}
\caption{(Colour online) The directional dependence of Young's moduli for \ce{{Cmc2}1-Si2P2X} (X=S, Se, Te and Po).} \label{fig-smp3}
\end{figure}

In figure~\ref{fig-smp3}, the 3D surface diagrams show an anisotropic degree of the Young's moduli, and the result shows that the Young's modulus of the \ce{{Cmc2}1-Si2P2Se} crystal is more anisotropic than the other compounds. For the \ce{{Cmc2}1-Si2P2X} compounds, according to the mechanical properties, the longitudinal sound velocity, transverse sound velocity and the density are calculated. They are shown as table~\ref{tbl-smp5}. The values of the density increase with increasing atomic number from S, Se, Te to Po. Based on the relationship between density and sound velocity, obviously, the longitudinal, transverse and average sound velocity all decrease with increasing atomic number from S, Se, Te to Po.

\subsection{Band structure and density of states }
\begin{figure}[!t]
\centering
\subfigure{
\includegraphics[width=6.8cm,height=6.8cm]{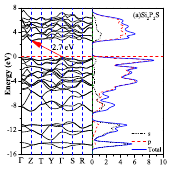}
}
\quad
\subfigure{
\includegraphics[width=6.8cm,height=6.8cm]{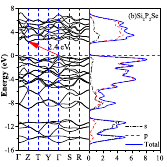}
}
\quad
\subfigure{
\includegraphics[width=6.8cm,height=6.8cm]{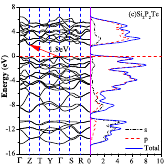}
}
\quad
\subfigure{
\includegraphics[width=6.8cm,height=6.8cm]{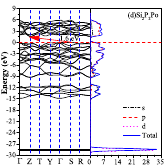}
}
\caption{(Colour online) Calculated band structures and total DOS for \ce{Si2P2X} (X=S, Se, Te and Po) using HSE06 at 0 GPa. The energy of the top of the valence band is set to zero.} \label{fig-smp4}
\end{figure}

\begin{figure}[H]
\centering
\subfigure{
\includegraphics[width=6.5cm,height=6.5cm]{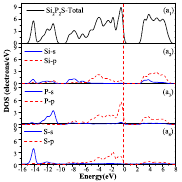}
}
\quad
\subfigure{
\includegraphics[width=6.5cm,height=6.5cm]{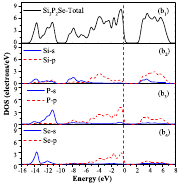}
}
\quad
\subfigure{
\includegraphics[width=6.5cm,height=6.5cm]{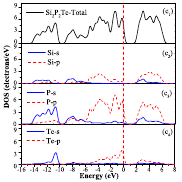}
}
\quad
\subfigure{
\includegraphics[width=6.5cm,height=6.5cm]{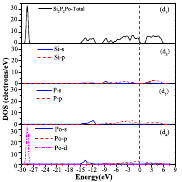}
}
\caption{(Colour online) Total and partial DOS (a) \ce{Si2P2S}, (b) \ce{Si2P2Se}, (c) \ce{Si2P2Te} and (d) \ce{Si2P2Po}.} \label{fig-smp5}
\end{figure}

The electronic properties of \ce{{Cmc2}1-Si2P2X} are analyzed using HSE06 at 0 GPa. The band structures are shown as figure~\ref{fig-smp4}. The top of the valence band is used as the reference and is expressed by a red dot line. The total and partial densities of states (DOS) are calculated and presented in figure~\ref{fig-smp5}. 

For \ce{{Cmc2}1-Si2P2S}, in figure~\ref{fig-smp4}~(a), the valence band maximum is at the highly symmetric $\Gamma$-point and the conduction band minimum is at the Z-point, which indicates that \ce{{Cmc2}1-Si2P2S} is an indirect bandgap semiconductor with the band gap of 2.7~eV. Figure~\ref{fig-smp5}~(a) shows the electronic DOS of \ce{{Cmc2}1-Si2P2S} in the energy range of $-16$ to 8~eV. The total DOS is mainly contributed by S-s in the range of $-15$ to $-13$~eV with some contribution from Si-s. Between $-13$ and $-11$~eV, the total DOS is mostly composed of P-s and a small fraction comes from Si-s and Si-p. In the range of $-9$ to $0$~eV, the total DOS is mainly derived from Si-p, P-p, S-p states, and a small part comes from Si-s. From 3 to 8 eV, the total DOS mainly includes three parts: Si-p, P-p and Si-s. 

Similarly, the \ce{{Cmc2}1-Si2P2Se} is also an indirect bandgap semiconductor with the band gap of 2.4~eV. From figure~\ref{fig-smp5}~(b), in the range of $-16$ to 8~eV, the trend of DOS for \ce{{Cmc2}1-Si2P2Se} is similar to that for \ce{{Cmc2}1-Si2P2S}. 

In figure~\ref{fig-smp4}~(c), \ce{{Cmc2}1-Si2P2Te} is an indirect bandgap semiconductor, whose bandgap is 1.8~eV. Figure~\ref{fig-smp5}~(c) shows that the electronic DOS of \ce{{Cmc2}1-Si2P2Te} is in the range from $-16$ to 8~eV. It can be noted that the total DOS in $-15$ to $-10$~eV is mainly from P-s, with some contribution of Si-s and Te-s. In the range of $-9$~eV to 0~eV, the total DOS is mainly composed of Si-p, P-p, Te-p states, and a small part comes from Si-s and P-s. In addition, the Si-p and P-p provide the main contribution in the range of 2 to 8~eV, and the P-s also plays an important role. 

From figure~\ref{fig-smp4}~(d), it can be concluded that the maximum value of the valence band of \ce{{Cmc2}1-Si2P2} Po is at the highly symmetric S-point, the conduction band minimum is at the Z-point. The indirect bandgap of \ce{{Cmc2}1-Si2P2Po} is 1.6~eV. In figure~\ref{fig-smp5}~(d), the total DOS situated in the range $-30$ to $-27$~eV has one peak value, which is mainly due to the Po-d state. In the range of $-15$ to $-12$~eV, the total DOS is mainly composed of P-s and Po-s states. Within the range of $-9$ to 0~ eV, the total DOS is mainly derived from contributions of the Si-p, P-p and Si-s. In the range of 1 to 6~eV, the projected density of states is mainly composed of Si-p and P-p, and a small part comes from the Si-s.

Obviously, for the \ce{{Cmc2}1-Si2P2X} compounds, the values of the band gaps decrease with increasing atomic number from S, Se, Te to Po.

\subsection{Optical properties}
The essential optical properties of materials, such as the dielectric function, refractive index, absorption, reflectivity, conductivity and loss function can be calculated.

\begin{figure}[!t]
\centering
\subfigure{
\includegraphics[width=6cm,height=5.5cm]{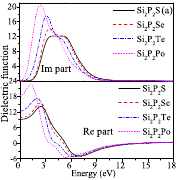}
}
\quad
\subfigure{
\includegraphics[width=6.1cm,height=5.6cm]{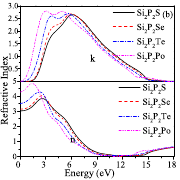}
}
\subfigure{
\includegraphics[width=6cm,height=5.6cm]{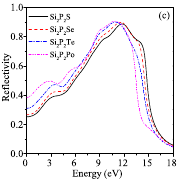}
}
\quad
\subfigure{
\includegraphics[width=6.1cm,height=5.5cm]{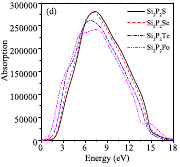}
}
\subfigure{
\includegraphics[width=6cm,height=6cm]{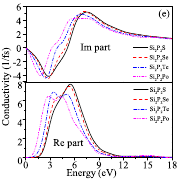}
}
\quad
\subfigure{
\includegraphics[width=6cm,height=6cm]{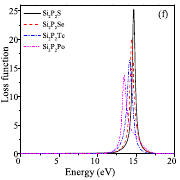}
}
\caption{(Colour online) Calculated optical properties (a) dielectric function, (b) refractive index, (c)~optical reflectivity spectrum, (d) absorption, (e) conductivity, (f) loss function.} \label{fig-smp6}
\end{figure}

For  \ce{Si2P2S},  \ce{Si2P2Se},  \ce{Si2P2Te} and  \ce{Si2P2Po}, the optical properties are calculated and presented in figure~\ref{fig-smp6}. The dielectric function can characterize how the materials respond to the incident electromagnetic wave. The calculated static dielectric functions are about 9.1, 9.8, 12.3 and 18.3, respectively. The real part curves of dielectric functions increase with increasing photon energy and reach the peak values at about 2.4, 2.3, 2.1 and 1.2~eV, respectively. After the maximum, the real part curves decrease with an increase of photon energy and then slowly and gradually increase close to zero. For \ce{Si2P2Se}, \ce{Si2P2Te} and \ce{Si2P2Po}, with an increase of photon energy, the curves of the imaginary part increase and attain the peak values at about 3.5, 3.1 and 2.4~eV, respectively. The values of the imaginary part approach zero when the energy exceeds 12 eV. In the range of 3.7 to 5.2~eV, the imaginary part of \ce{Si2P2S} reaches the highest value 12.1 and remains almost unchanged.

Figure~\ref{fig-smp6}~(b) shows the refractive index, $n(\omega)$, and extinction coefficient, $k(\omega)$. The static refractive indices $n(0)$ are found to be 3.0, 3.1, 3.5 and 4.3, respectively. Each $n(\omega)$ has a peak. The peaks are located at 2.6, 2.5, 2.3 and 1.4~eV, respectively. Each $k(\omega)$ value increases with an increasing photon energy, and then decreases after reaching the maximum. The $k(\omega)$ of \ce{{Cmc2}1-Si2P2S} and \ce{{Cmc2}1-Si2P2Se} have one peak, respectively. In both \ce{{Cmc2}1-Si2P2Te} and \ce{{Cmc2}1-Si2P2Po}, $k(\omega)$ has two main peaks. 

Figure~\ref{fig-smp6}~(c) shows the reflectivity of \ce{{Cmc2}1-Si2P2X}. The initial values are $25.3\%, 26.7\%, 31\%$ and $38.5\%$, respectively. Each reflectivity has a maximum value. These maximum values are located at 11.8, 11.6, 10.9 and 10.9~eV, respectively. 

The absorption spectra of \ce{{Cmc2}1-Si2P2X} are shown in figure~\ref{fig-smp6}~(d). We can notice that the absorption is basically located between 0 and 18~eV. For \ce{{Cmc2}1-Si2P2X}(S, Se, Te), the peaks of the absorption spectra are located at 7.3, 7.1 and 6.7~eV, respectively. For \ce{{Cmc2}1-Si2P2Po}, there are two remarkable peaks at 7.5 eV and 30.4~eV. 

The curves of the conductivities for \ce{{Cmc2}1-Si2P2X} are plotted in figure~\ref{fig-smp6}~(e). The real parts of the conductivity of these four structures are greater than zero because of electron excitation. The electrons in the valence band absorb the photon energy and the transition from the valence band to the conduction band takes place producing electron-hole pairs, then generating photoelectric signals. With an increase of photon energy, the real parts of the conductivities increase to the peak values and then decrease rapidly. The real part curves of \ce{Si2P2Te} and \ce{Si2P2Po} have two peaks each. For \ce{Si2P2S} and \ce{Si2P2Se}, the curves have one peak each. From figure~\ref{fig-smp6}~(e), the imaginary part values of the conductivities begin to be negative and decrease to their respective minima with increasing photon energy, then increase to respective positive peaks. After the maxima, the imaginary part curves decrease and gradually tend close to zero.

In figure~\ref{fig-smp6}~(f), the energy-loss spectra of \ce{{Cmc2}1-Si2P2X} are presented. The loss function describes the energy loss when an electron traverses the materials. The peaks of loss function represent the characteristic connected with the plasma oscillation. The corresponding frequency is the plasma frequency, above which the material exhibits a dielectric behavior [$\varepsilon_1(\omega)>0$], while below this frequency the material is characterized by metallic property [$\varepsilon_1(\omega)<0$]. For \ce{{Cmc2}1-Si2P2X}, the peaks of loss functions lie at 14.7, 14.5, 14.2 and 13.5~eV, respectively. Each peak corresponds to the sudden drop in its reflectivity. 

\section{Conclusion}
For the predicted new compounds, \ce{{Cmc2}1-Si2P2X} (X=S, Se, Te and Po), the orthorhombic structures are mechanically and dynamically stable, which are indicated by the calculations of the elastic constants and photon dispersion spectra. Their structural, electronic, mechanical and optical properties are investigated using the first principles. The results indicate that the bulk modulus of the \ce{{Cmc2}1-Si2P2Po} is larger than the others, which shows that it has a lower compressibility. The shear modulus of \ce{{Cmc2}1-Si2P2S} is the largest among \ce{{Cmc2}1-Si2P2X}, hence, \ce{{Cmc2}1-Si2P2S} has a better capability of resisting to the shape change. Poisson's ratio shows that \ce{{Cmc2}1-Si2P2S} has a better brittleness. The 3D maps of Young's modulus indicates that \ce{{Cmc2}1-Si2P2Se} is more anisotropic than the others. The maximum value of the valence band and the minimum value of the conduction band of the four compounds correspond to different K values in K space. These indicate that they are indirect-bandgap semiconductors. The values of the band gaps decrease with increasing `X' element atomic number, whereas the values of the calculated static dielectric constants and the static refractive indices increase with increasing atomic number from S, Se, Te, to Po. 

\section{Acknowledgements}
This work is supported by Natural Science Basic Research Program of Shaanxi (Program No.~2021JM-127). Part of calculations was done by the HPC system of Xidian University. We are grateful for the 111 Project (B17035).

\ukrainianpart

\title{Дослідження структури і фізичних властивостей \ce{{Cmc2}1-Si2P2X} за допомогою першопринципних розрахунків }

\author{Р. Янг\orcid{0000-0002-5962-5844}\refaddr{label1},
	С. Гао\refaddr{label1}, Ф. Ву\refaddr{label1}, К. Вей\refaddr{label1}, М. Сю\refaddr{label2}}
\addresses{
	\addr{label1} Школа фізики та оптоелектронної інженерії, Ксідіанський університет, Сіан, Шеньсі 710071, КНР
	\addr{label2} Коледж хімії та хімічної інженерії, Гуансійський університет, Наньнин, Гуансі 530004, КНР
}

\makeukrtitle

\begin{abstract}
	
		За допомогою першопринципних розрахунків з використанням методу фунціоналу густини передбачено існування нових структур \ce{{Cmc2}1-Si2P2X} (X=S, Se, Te і Po) та досліджено їх механічні, електронні та оптичні властивості. Пружні характеристики цих чотирьох сполук обчислено  методом ``напруження-деформація''. З розрахованих умов пружної стійкості та спектрів фононної дисперсії випливає, що ці сполуки є механічно і динамічно стійкими при нульовому тиску. Механічні характеристики, такі як модулі зсуву $G$, об'ємної пружності $B$, Юнга $E$ та коефіцієнти Пуассона $v$, розраховано в підході Фойгта-Рейса-Гілла.  Найбільшу твердість має \ce{{Cmc2}1-Si2P2S} завдяки найбільшому серед чотирьох сполук модулю Юнга, причому це ковалентний кристал. Також проаналізовано анізотропію їх механічних властивостей. Із розрахованих з використанням методу HSE06 зонних структур та густин станів випливає, що сполуки \ce{{Cmc2}1-Si2P2X} є напівпровідниками з непрямою забороненою зоною, а її ширина зменшується при збільшенні атомного числа від S, Se, Te, до Po. Окрім того, досліджено швидкості поздовжньої та поперечної звукової хвилі для  \ce{{Cmc2}1-Si2P2X}. Проаналізовано діелектричну сталу, втрату енергії електрона, показник заломлення, коефіцієнти відбивання, 
	поглинання та  провідність для оцінки оптичних характеристик \ce{Si2P2X}.

	\keywords сполуки \ce{{Cmc2}1-Si2P2X} (X=S, Se, Te, та Po), перші принципи, механічні властивості, електронні властивості, оптичні властивості
	
\end{abstract}

\lastpage
\end{document}